\documentclass[12pt]{article}
\begin{document}

\def\ds{\displaystyle}
\def\beq{\begin{equation}}
\def\eeq{\end{equation}}
\def\bea{\begin{eqnarray}}
\def\eea{\end{eqnarray}}
\def\beeq{\begin{eqnarray}}
\def\eeeq{\end{eqnarray}}
\def\ve{\vert}
\def\vel{\left|}
\def\ver{\right|}
\def\nnb{\nonumber}
\def\ga{\left(}
\def\dr{\right)}
\def\aga{\left\{}
\def\adr{\right\}}
\def\lla{\left<}
\def\rra{\right>}
\def\rar{\rightarrow}
\def\nnb{\nonumber}
\def\la{\langle}
\def\ra{\rangle}
\def\ba{\begin{array}}
\def\ea{\end{array}}
\def\tr{\mbox{Tr}}
\def\ssp{{\Sigma^{*+}}}
\def\sso{{\Sigma^{*0}}}
\def\ssm{{\Sigma^{*-}}}
\def\xis0{{\Xi^{*0}}}
\def\xism{{\Xi^{*-}}}
\def\qs{\la \bar s s \ra}
\def\qu{\la \bar u u \ra}
\def\qd{\la \bar d d \ra}
\def\qq{\la \bar q q \ra}
\def\gGgG{\la g^2 G^2 \ra}
\def\q{\gamma_5 \not\!q}
\def\x{\gamma_5 \not\!x}
\def\g5{\gamma_5}
\def\sb{S_Q^{cf}}
\def\sd{S_d^{be}}
\def\su{S_u^{ad}}
\def\sbp{{S}_Q^{'cf}}
\def\sdp{{S}_d^{'be}}
\def\sup{{S}_u^{'ad}}
\def\ssp{{S}_s^{'??}}

\def\sig{\sigma_{\mu \nu} \gamma_5 p^\mu q^\nu}
\def\fo{f_0(\frac{s_0}{M^2})}
\def\ffi{f_1(\frac{s_0}{M^2})}
\def\fii{f_2(\frac{s_0}{M^2})}
\def\O{{\cal O}}
\def\sl{{\Sigma^0 \Lambda}}
\def\es{\!\!\! &=& \!\!\!}
\def\ap{\!\!\! &\approx& \!\!\!}
\def\ar{&+& \!\!\!}
\def\ek{&-& \!\!\!}
\def\kek{\!\!\!&-& \!\!\!}
\def\cp{&\times& \!\!\!}
\def\se{\!\!\! &\simeq& \!\!\!}
\def\eqv{&\equiv& \!\!\!}
\def\kpm{&\pm& \!\!\!}
\def\kmp{&\mp& \!\!\!}
\def\mcdot{\!\cdot\!}


\def\simlt{\stackrel{<}{{}_\sim}}
\def\simgt{\stackrel{>}{{}_\sim}}


\title{
         {\Large
                 {\bf
Nucleon form factors induced by isovector and isoscalar
axial--vector currents in QCD
                 }
         }
      }

\author{\vspace{1cm}\\
{\small T. M. Aliev \thanks
{e-mail: taliev@metu.edu.tr}~\footnote{permanent address:Institute
of Physics,Baku,Azerbaijan}\,\,,
M. Savc{\i} \thanks
{e-mail: savci@metu.edu.tr}} \\
{\small Physics Department, Middle East Technical University,
06531 Ankara, Turkey} }

\date{}

\begin{titlepage}
\maketitle
\thispagestyle{empty}

\begin{abstract}
Using the most general form of the baryon current, nucleon form factors, 
induced by isovector and isoscalar axial--vector currents, are studied in 
the framework of light cone QCD sum rule approach. Comparison of
our results on form factors with the existing results and lattice
calculations are presented.  
\end{abstract}

~~~PACS number(s): 11.55.Hx, 13.40.Em, 14.20.Jn
\end{titlepage}

\section{Introduction}

It is well known that the structure of the nucleon is parametrized in 
terms of form factors. The electromagnetic form factors of nucleon are 
measured in a wide range of momentum transfer squared $q^2$ (see for 
example \cite{R8301} and the references therein). But, in contrast to 
the electromagnetic case, the form
factors $G_A(q^2)$, $G_P(q^2)$ and $G_T(q^2)$ due to isovector axial--vector 
current are not known. The nucleon matrix elements of axial--vector currents
at $q^2=0$ are determined by the axial--vector coupling constants such as 
$g_A$ (isovector), $g_A^8$ (octet), $g_A^S$ (isoscalar) and $g_A^0$
(flavor singlet).
Among these only nucleon isovector coupling constant 
$g_A$ is well known which is measured from the neutron $\beta$--decay.
Knowledge of any three of these coupling constants determines the quark spin
content of the nucleon. For this reason study of these constants receives
great interest.

Using the Lorentz covariance, the matrix element of the isovector axial--vector
current between initial and final states is parametrized as
\bea
\label{e8301}
\lla N(p^\prime) \vel A_\mu \ver N(p) \rra = \bar{u}(p^\prime) \Bigg[
\gamma_\mu \gamma_5 G_A(q^2) + \frac{q_\mu}{2 m_N} \gamma_5
G_P(q^2) + i \sigma_{\mu\nu} \frac{q^\nu}{2 m_N} \gamma_5 G_T(q^2)\Bigg] u(p)~,
\eea    
where $A_\mu = \bar{u}\gamma_\mu\gamma_5 u - \bar{d}\gamma_\mu\gamma_5 d$,
$q=p-p^\prime$, $m_N$ is the nucleon mass, 
and $G_A$, $G_P$ and $G_T$ are the axial, induced pseudoscalar and induced
tensorial form factors, respectively, induced 
by the isovector and axial--vector currents. 
The matrix element of isoscalar axial--vector current
$A_\mu^S=\bar{u}\gamma_\mu\gamma_5 u + \bar{d}\gamma_\mu\gamma_5 d$ 
between nucleon
states is determined similar to (\ref{e8301}) with the following
replacements: $G_A \rar G_A^S$, $G_P \rar G_P^S$ and 
$G_T \rar G_T^S$ (here the superscript $S$ means isoscalar). The form
factors $G_T$ and $G_T^S$ both vanish as a result of the exact
isospin symmetry and G--parity invariance of the strong interaction.

The aim of the present work is the calculation of the form factors
$G_A(G_A^S)$ and $G_P(G_P^S)$ in the framework of light cone QCD sum
rules (LCSR) \cite{R8303,R8304} approach, using the most general form of 
nucleon interpolating current. LCSR is based on the operator product
expansion over twist of the operators near the light cone. This method combines
the standard sum rules technique \cite{R8305} with the parton distribution
amplitudes describing the hard exclusive processes. This method is widely
applied to the problems in the meson sector (see \cite{R8304}). Lately, the
electromagnetic form factors of nucleon \cite{R8306}, the scalar form factor
of nucleon \cite{R8307} and the weak $\Lambda_b \rar p \ell \nu$
\cite{R8308} are investigated in the baryonic sector in the framework of this 
method. Note that higher twist amplitudes of the nucleon are calculated in
\cite{R8309}.

The outline of this work is as follows. In section 2 the basic ingredients
for calculating the form factors due to axial--vector current are introduced
and sum rules for the form factors are constructed. In section 3 we present
our numerical results together with the concluding remarks.   

\section{Sum rules for the form factors of the nucleon due to isovector and
isoscalar currents}

In this section, we construct sum rules for the form factors of the
nucleon due to the isovector and isoscalar axial--vector currents. For this
purpose we start by considering the polarization operator, which is the basic
object of the sum rule approach,
\bea
\label{e8302}
\Pi_\mu (p,q) = i \int d^4x e^{iqx} \lla 0 \vel T\{\eta(0) A_\mu^{(S)} (x)
\} \ver N(p) \rra~,
\eea
where
\bea
A_\mu^{(S)} = \bar{u} \gamma_\mu \gamma_5 u \mp 
\bar{d} \gamma_\mu \gamma_5 d~, \nnb
\eea
with the upper (lower) sign corresponding to the isovector (isoscalar) 
axial--vector current, and $\eta$ is an interpolating current with nucleon quantum
numbers. The nucleon interpolating currents without the derivative terms,
and with nucleon quantum numbers, can be written as \cite{R8310}
\bea
\label{e8303}
\eta_1 (x) \es 2 \epsilon^{abc} \sum_{\ell=1}^2 \Big( u^{aT}(x) C
A_1^\ell d^b (x) \Big) A_2^\ell u^c (x) ~, \\
\label{e8304}
\eta_2 (x) \es \frac{2}{3} \epsilon^{abc} \Big[ \Big( u^{aT}(x) C
\rlap/z u^b (x) \Big) \gamma_5 \rlap/z d^c (x)- 
\Big( u^{aT}(x) C \rlap/z d^b (x) \Big) \gamma_5 \rlap/z u^c (x) \Big]~,
\eea 
where $A_1^1=I$, $A_2^1=\gamma_5$ , $A_1^2=\gamma_5$ , $A_2^2=\beta$; $C$
is the charge conjugation operator; $a$, $b$, $c$ are the color indices, 
and $z$ is a light--like vector with $z^2=0$. The choice $\beta=-1$ in Eq.
(\ref{e8303}) corresponds to the Ioffe current \cite{R8311}. Note that
$\eta_2$ current is modified in the following way \cite{R8309}
\bea
\label{e8305}
\eta_3 (x) = \epsilon^{abc} \Big[ \Big( u^{aT}(x) C \rlap/z
u^b (x) \Big) \gamma_5 \rlap/z d^c (x) \Big]~.
\eea
The axial form factor and induced pseudoscalar form factor of the
nucleon are calculated in \cite{R8312} in the framework of LCSR using the
current $\eta_3(x)$. However the current $\eta_3(x)$ couples to both spin
$1/2$ and $3/2$ baryons. Therefore, for a reliable determination of the form
factors the unwanted contributions coming from spin $3/2$ states should be
eliminated. But this elimination is not done in \cite{R8312}.

Note that the form factors $G_A(Q^2)$ and $G_P(Q^2)$ induced by the
isovector axial current are calculated in \cite{R8313} in LCSR using the
Ioffe current.  

In the present work we calculate the form factors induced by isovector and
isoscalar axial--vector currents using the general form of the interpolating 
$\eta_1(x)$.

We start by calculating the correlation function (\ref{e8301}) from the QCD
side. At large Euclidean momenta $p^{\prime 2} = (p-q)^2$ and $q^2=-Q^2$ the
correlation function can be calculated perturbatively. Using the expressions
of interpolating currents we get the following result for the correlator:
\bea
\label{e8306}
\Pi_\mu \es \frac{1}{2} \int d^4x e^{iqx} \sum_{\ell=1}^2 \Bigg\{
\ga C A_1^\ell\dr_{\alpha\gamma} \Big[ A_2^\ell S_u(-x) \gamma_\mu 
\gamma_5\Big]_{\rho\beta}
4 \epsilon^{abc} \lla 0 \vel u_\alpha^a(0) u_\beta^b(x)
d_\gamma^c(0) \ver N \rra \nnb \\ 
\ar \ga A_2^\ell \dr_{\rho\alpha} \Big[ \ga C A_1^\ell\dr^T S_u(-x)
\gamma_\mu\gamma_5\Big]_{\gamma\beta} 4 \epsilon^{abc} \lla 0 \vel
u_\alpha^a(x)
u_\beta^b(x) d_\gamma^c(0) \ver N \rra \nnb \\
\kmp \ga A_2^\ell \dr_{\rho\beta} \Big[ C A_1^\ell S_u(-x)
\gamma_\mu\gamma_5\Big]_{\alpha\gamma} 4 \epsilon^{abc} \lla 0 \vel u_\alpha^a(0)
u_\beta^b(0)  d_\gamma^c(x) \ver N \rra \Bigg\}~,
\eea
where $S_q(-x)$ is the light quark propagator and its light cone expanded
expression is \cite{R8313}
\bea
\label{e8307}
S(x) = \frac{i \rlap/x}{2 \pi^2 x^4} - \frac{\la q \bar{q} \ra}{12} \Bigg(
1+\frac{m_0^2 x^2}{16} \Bigg)
- i g_s \int_0^1 dv \Bigg[ \frac{\rlap/x}{16 \pi^2 x^4} G_{\mu\nu}
\sigma^{\mu\nu} - v x^\mu G_{\mu\nu} \gamma^\nu \frac{i}{4 \pi^2 x^2}
\Bigg]~.
\eea
The terms proportional to $G_{\mu\nu}$ in Eq. (\ref{e8307}) give
contribution to four-- and five--particle nucleon distribution functions, and
these amplitudes are expected to be small \cite{R8314,R8315} which will be
neglected in further analysis. Only the first term in Eq. (\ref{e8307})
survives since Borel transformation kills the second term.

It follows from Eq. (\ref{e8306}) that for the calculation of $\Pi_\mu$ we
need to know the matrix element $4 \epsilon^{abc} \lla 0 \vel
u_\alpha^a(a_1x) u_\beta^b(a_2x) d_\gamma^c(a_3x) \ver N(p) \rra$. This
matrix element of nonlocal operator is defined in terms of the nucleon
distribution amplitudes (DAs), and their explicit expressions are presented 
in \cite{R8309,R8313,R8316,R8317,R8318}.

Using the explicit expressions of the nucleon DAs and performing integration
over $x$ and selecting the structures $\rlap/q \gamma_\mu \gamma_5$ for
$G_A(G_A^S)$ and $q_\mu \rlap/q \gamma_5$ for $G_P(G_P^S)$, we get

\bea
\label{e8308}
-{\lambda_N \over m_N^2 - p^{\prime 2}} G_A \es {1\over 2} \Bigg\{
m_N \int_0^1 {dt_2 \over (q-pt_2)^2} \Big[ (1-\beta) F_1(t_2) +
(1+\beta) F_2(t_2) \Big] \nnb \\
\kmp {m_N \over 2} \int_0^1 {dt_3 \over (q-pt_3)^2} \Big[ 2 (1-\beta) F_3(t_3) +
(1+\beta) F_4(t_3) \Big] \nnb \\
\ar m_N^3 \int_0^1 {dt_2 \over (q-pt_2)^4} \Big[ (1-\beta) F_5(t_2) +
(1+\beta) F_6(t_2) \Big] \nnb \\
\kmp m_N^3 \int_0^1 {dt_3 \over (q-pt_3)^4} \Big[ (1-\beta) F_7(t_3) +
(1+\beta) F_8(t_3) \Big] \nnb \\
\ar m_N^3 \int_0^1 {dt_2 \over (q-pt_2)^4} \Big[ (1-\beta) F_9(t_2) +
(1+\beta) F_{10}(t_2) \Big] \nnb \\
\kmp m_N^3 \int_0^1 {dt_3 \over (q-pt_3)^4} \Big[ (1-\beta) F_{11}(t_3) +
(1+\beta) F_{12}(t_3) \Big] \Bigg\}~, \\ \nnb \\
\label{e8309}
-{\lambda_N \over m_N^2 - p^{\prime 2}} G_P \es {1\over 2} \Bigg\{
m_N^2 \int_0^1 {dt_2 \over (q-pt_2)^4} \Big[ 2 (1-\beta)
F_{13}(t_2) + (1+\beta) F_{14}(t_2) \Big] \nnb \\
\kmp m_N^2 \int_0^1 {dt_3 \over (q-pt_3)^4} \Big[ 2 (1-\beta)
F_{15}(t_3) + (1+\beta) F_{16}(t_3) \Big] \Bigg\}~,
\eea
where upper (lower) sign corresponds to isovector (isoscalar) axial--vector
current, and
\bea
F_1(t_2) \es \int_0^{1-t_2} dt_1\Big[ - \widetilde{\cal A}_2 - 2 A_3 -
\widetilde{\cal V}_2 + 2 V_1 - 2 V_3 \Big](t_1,t_2,1-t_1-t_2)~, \nnb \\
F_2(t_2) \es \int_0^{1-t_2} dt_1\Big[ 2 P_1 + 2 S_1 + 4 T_1 - \widetilde{\cal
T}_2 - 8 T_7 - 3 \widetilde{\cal T}_4  \Big] (t_1,t_2,1-t_1-t_2)~, \nnb \\
F_3(t_3) \es \int_0^{1-t_3} dt_1\Big[ A_1 - V_1 \Big](t_1,1-t_1-t_3,t_3)~, \nnb \\
F_4(t_3) \es \int_0^{1-t_3} dt_1\Big[ 2 P_1 + 2 S_1 - 2 T_1 + \widetilde{\cal T}_2 
+ 4 T_7 + \widetilde{\cal T}_4 \Big] (t_1,1-t_1-t_3,t_3)~, \nnb \\
F_5(t_2) \es \int_0^{1-t_2} dt_1 \Big[2 
\widetilde{\cal V}_1^M \Big](t_1,t_2,1-t_1-t_2)~, \nnb \\
F_6(t_2) \es \int_0^{1-t_2} dt_1\Big[ 
4\widetilde{\cal T}_1^M \Big](t_1,t_2,1-t_1-t_2)~, \nnb \\
F_7(t_3) \es \int_0^{1-t_3} dt_1\Big[ 
\widetilde{\cal A}_1^M - \widetilde{\cal V}_1^M\Big](t_1,1-t_1-t_3,t_3)~, \nnb \\
F_8(t_3) \es \int_0^{1-t_3} dt_1                    
\Big[-\widetilde{\cal T}_1^M\Big] (t_1,1-t_1-t_3,t_3)~, \nnb \\
F_9 (t_2) \es \int_1^{t_2} d\lambda \int_1^\lambda d\rho \int_0^{1-\rho} dt_1
\Big[ 2 \widetilde{\cal A}_6\Big] (t_1,\rho,1-t_1-\rho)~, \nnb \\
F_{10} (t_2) \es \int_1^{t_2} d\lambda \int_1^\lambda d\rho \int_0^{1-\rho} dt_1
\Big[ - 2 \widetilde{\cal T}_6 + 4 \widetilde{\cal T}_8\Big]
(t_1,\rho,1-t_1-\rho)~, \nnb \\
F_{11} (t_3) \es \int_1^{t_3} d\lambda \int_1^\lambda d\rho \int_0^{1-\rho} dt_1
\Big[ \widetilde{\cal A}_6 + \widetilde{\cal V}_6 \Big] (t_1,1-t_1-\rho,\rho)~, \nnb \\
F_{12} (t_3) \es \int_1^{t_3} d\lambda \int_1^\lambda d\rho \int_0^{1-\rho} dt_1
\Big[ \widetilde{\cal T}_6 - \widetilde{\cal T}_8 \Big]
(t_1,1-t_1-\rho,\rho)~, \nnb \\
F_{13} (t_2) \es \int_1^{t_2} d\rho \int_0^{1-\rho} dt_1 \Big[
\widetilde{\cal A}_2 - 2 \widetilde{\cal A}_4 + 2 \widetilde{\cal A}_5 +
\widetilde{\cal V}_2 - \widetilde{\cal V}_5 \Big] (t_1,\rho,1-t_1-\rho)~, \nnb \\
F_{14} (t_2) \es \int_1^{t_2} d\rho \int_0^{1-\rho} dt_1 \Big[
2 (\widetilde{\cal P}_2 - \widetilde{\cal S}_2) + 4 \widetilde{\cal T}_2 +
6 \widetilde{\cal T}_4 + 10 \widetilde{\cal T}_5 - \widetilde{\cal T}_6
+ 20 \widetilde{\cal T}_7 \Big] (t_1,\rho,1-t_1-\rho)~, \nnb \\
F_{15} (t_3) \es \int_1^{t_3} d\rho \int_0^{1-\rho} dt_1 \Big[
- \widetilde{\cal A}_2 - \widetilde{\cal A}_5 - \widetilde{\cal V}_2 +
\widetilde{\cal V}_5 \Big] (t_1,1-t_1-\rho,\rho)~, \nnb \\
F_{16} (t_3) \es \int_1^{t_3} d\rho \int_0^{1-\rho} dt_1 \Big[
2 (\widetilde{\cal P}_2 - \widetilde{\cal S}_2) + 2 \widetilde{\cal T}_4 +
2 \widetilde{\cal T}_5 - \widetilde{\cal T}_6 + 4 \widetilde{\cal T}_7
\Big] (t_1,1-t_1-\rho,\rho)~.\nnb
\eea
Those functions appearing in Eqs. (\ref{e8308}) and (\ref{e8309}) are determined as
\bea
\widetilde{\cal V}_2(t_i) \es V_1(t_i) - V_2(t_i) - V_3(t_i)~, \nnb \\
\widetilde{\cal A}_2(t_i) \es - A_1(t_i) + A_2(t_i) - A_3(t_i)~, \nnb \\
\widetilde{\cal A}_4(t_i) \es - 2 A_1(t_i) - A_3(t_i) - A_4(t_i)
+ 2 A_5(t_i)~, \nnb \\
\widetilde{\cal A}_5(t_i) \es A_3(t_i) - A_4(t_i)~, \nnb \\ 
\widetilde{\cal A}_6(t_i) \es A_1(t_i) - A_2(t_i) + A_3(t_i) +
A_4(t_i) - A_5(t_i) + A_6(t_i)~, \nnb \\
\widetilde{\cal T}_2(t_i) \es T_1(t_i) + T_2(t_i) - 2 T_3(t_i)~, \nnb \\
\widetilde{\cal T}_4(t_i) \es T_1(t_i) - T_2(t_i) - 2 T_7(t_i)~, \nnb \\
\widetilde{\cal T}_5(t_i) \es - T_1(t_i) + T_5(t_i) + 2 T_8(t_i)~, \nnb \\
\widetilde{\cal T}_6(t_i) \es 2 \Big[T_2(t_i) - T_3(t_i) - T_4(t_i) +
T_5(t_i) + T_7(t_i) + T_8(t_i)\Big]~, \nnb \\
\widetilde{\cal T}_7(t_i) \es T_7(t_i) - T_8(t_i)~, \nnb \\
\widetilde{\cal S}_2(t_i) \es S_1(t_i) - S_2(t_i)~, \nnb \\ 
\widetilde{\cal P}_2(t_i) \es P_2(t_i) - P_1(t_i)~, \nnb
\eea
whose explicit expressions are given in \cite{R8313}.

Physical part of the correlation function is obtained by
inserting a complete set of states between the currents in Eq. (\ref{e8301})
with the same quantum numbers of the current $\eta(x)$. After isolating 
the pole term of the nucleon state, the correlator function (\ref{e8301})
can be written as    
\bea
\label{e8310}
\Pi_\mu \es \frac{\lla 0 \vel \eta \ver N(p^\prime) \rra
\lla N(p^\prime) \vel A_\mu^S \ver N(p) \rra}
{m_N^2-p^{\prime 2}} +
\sum_{h} \frac{\lla 0 \vel \eta \ver h(p^\prime) \rra
\lla h(p^\prime) \vel A_\mu^S \ver N (p) \rra}
{m_{h}^2-p^{\prime 2}}~,
\eea
where $p^\prime=p-q$ and $q$ is the momentum carried by the axial--vector
current. The second term in (\ref{e8310}) describes the higher states and
continuum contribution, and $h$ is complete set of the hadrons with the
quantum numbers of the ground state nucleon.

Contribution of higher states to the physical part of the sum rules are
taken into account using quark--hadron duality, i.e., spectral density for
higher states is equal to the perturbative spectral density starting from
$s>s_0$, where $s_0$ is the continuum threshold.

The matrix elements entering to Eq. (\ref{e8310}) are defined as
\bea
\label{e8311}
\lla 0 \vel \eta \ver N(p^\prime) \rra \es \lambda_N u_N(p^\prime)~,\\
\label{e8312}
\lla N(p^\prime) \vel A_\mu^S \ver N(p) \rra \es 
\bar{u}_N(p^\prime) \Bigg[ \gamma_\mu G_A^S(q^2) + \frac{q_\mu}{2 m_N}
G_P^S(q^2) \Bigg] \gamma_5 u_N(p)~.
\eea
Substituting Eqs. (\ref{e8311}) and (\ref{e8312}) into Eq. (\ref{e8310}) and selecting
the structures $\rlap/q\gamma_\mu\gamma_5$ and $q_\mu\rlap/q\gamma_5$
and performing Borel transformation with respect to the variable $(q-p)^2$,
which suppresses the continuum and higher state contributions, we get the
above--mentioned form factors:
\bea
\label{e8313}
G_A \es - {1 \over 2 \lambda_N} e^{m_N^2/M^2} \Bigg\{
-m_N \int_{x_0}^1 dt_2 e^{-s(t_2)/M^2} \Big[ (1-\beta) 
F_1(t_2) + (1+\beta) F_2(t_2) \Big] \nnb \\
\kmp \Bigg( - {m_N\over 2} \int_0^1 dt_3 e^{-s(t_3)/M^2} \Big[ 2 (1-\beta) 
F_3(t_3) + (1+\beta) F_4(t_3) \Big]\Bigg) \nnb \\
\ar {m_N^3 \over M^2} \int_{x_0}^1 {dt_2\over t_2^2}
e^{-s(t_2)/M^2} \Big[ (1-\beta) F_5(t_2) +
(1+\beta) F_6(t_2) \Big] \nnb \\
\ar {m_N^3 \over Q^2+x_0^2 m_N^2} e^{-s_0/M^2}
\Big[ (1-\beta) F_5(x_0) + (1+\beta) F_6(x_0) \Big] \nnb \\
\kmp {m_N^3 \over M^2} \int_{x_0}^1 {dt_3\over t_3^2}
e^{-s(t_3)/M^2} \Big[ (1-\beta) F_7(t_3) +
(1+\beta) F_8(t_3) \Big] \nnb \\
\kmp {m_N^3 \over Q^2+x_0^2 m_N^2} e^{-s_0/M^2}
\Big[ (1-\beta) F_7(x_0) + (1+\beta) F_8(x_0) \Big] \nnb \\
\ar {m_N^3 \over M^2} \int_{x_0}^1 {dt_2\over t_2^2}
e^{-s(t_2)/M^2} \Big[ (1-\beta) F_9(t_2) +
(1+\beta) F_{10}(t_2) \Big] \nnb \\
\ar {m_N^3 \over Q^2+x_0^2 m_N^2} e^{-s_0/M^2}
\Big[ (1-\beta) F_9(x_0) + (1+\beta) F_{10}(x_0) \Big] \nnb \\
\kmp {m_N^3 \over M^2} \int_{x_0}^1 {dt_3\over t_3^2}  
e^{-s(t_3)/M^2} \Big[ (1-\beta) F_{11}(t_3) +
(1+\beta) F_{12}(t_3) \Big] \nnb \\
\kmp {m_N^3 \over Q^2+x_0^2 m_N^2} e^{-s_0/M^2}
\Big[ (1-\beta) F_{11}(x_0) + (1+\beta) F_{12}(x_0) \Big] \Bigg\}~, \\ \nnb \\
\label{e8314}
G_P \es - {1 \over 2 \lambda_N} e^{m_N^2/M^2} \Bigg\{
{m_N^2 \over M^2} \int_{x_0}^1 dt_2 e^{-s(t_2)/M^2} \Big[ 2 (1-\beta)
F_{13}(t_2) + (1+\beta) F_{14}(t_2) \Big] \nnb \\
\ar {m_N^3 \over Q^2+x_0^2 m_N^2} e^{-s_0/M^2}
\Big[ 2 (1-\beta) F_{13}(x_0) + (1+\beta) F_{14}(x_0) \Big] \nnb \\
\kmp {m_N^2 \over M^2} \int_{x_0}^1 dt_3 e^{-s(t_3)/M^2} \Big[ 2
(1-\beta) F_{15}(t_3) + (1+\beta) F_{16}(t_3) \Big] \nnb \\
\kmp {m_N^3 \over Q^2+x_0^2 m_N^2} e^{-s_0/M^2}
\Big[ 2 (1-\beta) F_{15}(x_0) + (1+\beta) F_{16}(x_0) \Big] \Bigg\}~.
\eea

where
\bea
\lambda_N^2 \es e^{m_N^2/M^2} \Bigg\{\frac{M^6}{256 \pi^4} E_2(x) (5+2 \beta +
\beta^2) - (1-\beta^2) \frac{\la \bar{u}u \ra}{6} \Big[ 6 \la \bar{d}d \ra +
\la \bar{u}u \ra \Big] \nnb \\
\ar (1-\beta^2) \frac{m_0^2}{24 M^2}  \la \bar{u}u \ra \Big[12 \la \bar{d}d
\ra + \la \bar{u}u \ra \Big]\Bigg\}~, \nnb
\eea
and
\bea
E_2(s_0/M^2) = 1-e^{s_0/M^2} \sum_{k=0}^2 \frac{(s_0/M^2)^k}{k!}~. \nnb
\eea

In performing Borel transformation, we use the following substitution rules
(see for example \cite{R8312,R8313} and \cite{R8319})
\bea
\label{e8315}
\int dx \frac{\rho(x)}{(q-xp)^2} &\rar& - 
\int \frac{dx}{x}\rho(x) e^{-s/M^2}~, \nnb \\ 
\int dx \frac{\rho(x)}{(q-xp)^4} &\rar& \frac{1}{M^2}
\int \frac{dx}{x^2}\rho(x) e^{-s/M^2}
+ \frac{\rho(x_0)}{Q^2+x_0^2 m_N^2}e^{-s_0/M^2}~,
\eea
where 
\bea
s(x) \es (1-x) m_N^2 + \frac{1-x}{x} Q^2~, \nnb
\eea
and $x_0$ is the solution of the quadratic equation for $s=s_0$, i.e., 
\bea
\label{e8316}
x_0 \es \frac{1} {2 m_N^2}\Big[
\sqrt{(Q^2+s_0-m_N^2)^2 + 4 m_N^2 Q^2} - (Q^2+s_0-m_N^2)\Big]~, \nnb
\eea
and $Q^2 = - q^2$.

\section{Results and discussion}

Now we are ready to examine the sum rules for the form factors. It follows from 
expressions of the sum rules for form factors that DAs are the main input
parameters of LCSR. The complete list of all DAs which enter to the sum
rules for form factors can be found in \cite{R8313}.

These DAs contain eight hadronic parameters $f_N$, $\lambda_1$, $\lambda_2$,
$V_1^d$, $A_1^u$, $f_1^d$, $f_2^d$ and $f_1^u$ which should be determined in
the framework of various models. In further numerical calculations we
consider three different sets of these parameters:

\begin{itemize}

\item 
QCD sum rules based DAs, in which the parameters are determined from QCD
sum rules (set 1) having the values $A_1^u = 0.38 \mp 0.15$, $V_1^d = 0.23
\mp 0.03$, $f_1^d = 0.40 \mp 0.05$, $f_2^d = 0.22 \mp 0.05$, $f_1^u = 0.07
\mp 0.05$.

\item 
A model for nucleon DAs (set 2) in which the above--mentioned parameters 
are chosen in such a way that the form of nucleon DAs describe well the 
existing experimental data on nucleon form factors (see \cite{R8313}), 
whose numerical values are $A_1^u = 1/4$, $V_1^d = 13/42$ and the values 
of $f_1^d$, $f_2^d$ and  $f_1^u$ are the same as in set 1.

\item Asymptotic forms of DAs of all twists (set 3) in which the values 
of the parameters are set to $A_1^u = 0$, $V_1^d = 1/3$, $f_1^d = 3/10$, 
$f_2^d = 4/15$, $f_1^u = 1/10$.      

\end{itemize}

Note that the values of $f_N$, $\lambda_1$ and $\lambda_2$ for these three 
sets are the same, i.e., $f_N = (5.0 \mp 0.5)\times 10^{-3}~GeV^2$, 
$\lambda_1 = - (2.7 \mp 0.9)\times 10^{-2}~GeV^2$ and $\lambda_2 = (5.4
\mp 1.9)\times 10^{-2}~GeV^2$.  
  
The values of the non--perturbative parameters entering to DAs at 
$\mu=1~GeV$ scale
are given in \cite{R8309} and \cite{R8313}. Sum rules for the form factors
involve three unphysical parameters, namely, continuum threshold $s_0$,
Borel parameter $M^2$ and the parameter $\beta$ in the interpolating current of
nucleon. Of course, if we can perform OPE up to infinite order, the
result must be independent of these parameters. But we truncated OPE in the
finite order, and hence, there appears dependence of sum rules on these
auxiliary parameters. However, any physical quantity can not depend on
unphysical parameters. Therefore, in the first hand, we should find the
appropriate regions of unphysical parameters where form factors are
independent on them. From an analysis of mass sum rules it follows that,
when continuum threshold $s_0$ lies in the region $2~GeV^2 \le s_0 \le
2.5~GeV^2$, the prediction of sum rules on the mass of the baryons are
practically independent of $s_0$. It is this region of $s_0$ which we will
use in our numerical calculations.

Having determined $s_0$, we next
try to find a  region (the so called working region) of $M^2$ where 
the above--mentioned form factors are independent of $M^2$ at fixed 
values of $s_0$ and $\beta$. We study the dependence of
the form factors induced by the axial--vector current on $M^2$ at
fixed values of $Q^2$ and $\beta$ at $s_0=2.0~GeV^2$ and $s_0=2.5~GeV^2$, 
for three different sets of the wave functions. We obtain that the results 
are almost the same for both choices of $s_0$.

An upper bound for the Borel parameter $M^2$ is determined by requiring that
the contribution of continuum be less compared to the continuum subtracted sum
rules. Lower limit is determined from the condition that the contribution of
the term with highest power of $1/M^2$ is less, say 30\%, compared to the
higher powers of $M^2$ term contribution. Using these constraints we found
that the working region of $M^2$ is $1~GeV^2 \le M^2 \le 2.5~GeV^2$, and
we obtain that the results are rather stable with respect to
the variations of $M^2$, when $M^2$ lies in the above--mentioned working
region. It is also observed that the results of for the form factors depend
also on the parameter $\beta$. But, as has already been mentioned, the
parameter $\beta$ is an auxiliary quantity and therefore the form factors
must be independent of it. As a result of this argument, we need to find a
region for $\beta$, where the results for the form factors are independent
of its value. Analyses of mass sum rules \cite{R8319} and meson--octet baryon 
couplings \cite{R8320} lead to the result $\beta<-1.3$ and $\beta>3.3$. 
In our numerical calculations we will use these bounds
for $\beta$. It should be noted here that the Monte Carlo analysis of mass
sum rules for baryons \cite{R8321} predicts an optimal value $\beta\simeq
-1.2$, which is close to our choice of lower bound.

The correlation functions can be calculated in QCD for sufficiently large
negative values of $Q^2$ and $(p-q)^2$ using OPE. The form factors can
reliably be determined at the range $Q^2 \ge 2~GeV^2$. Our approach is not 
applicable for smaller values of $Q^2$. For this reason, the form factors
are evaluated in the range $Q^2 \ge 2~GeV^2$.
We study the dependence of the form factors on $Q^2$ at fixed
values of $\beta$, lying within the above--mentioned working region, at
fixed values of $M^2=2~GeV^2$ and $s_0=2.25~GeV^2$, for the three sets of DAs.

From numerical analysis we obtain the following results:

\begin{itemize}

\item The form factors $G_A(Q^2)$ and $G_P(Q^2)$ exhibit practically the
same $Q^2$ dependence for all three sets of the nucleon DAs. 

\item The values of the form factor $G_A(Q^2)$ almost coincide 
at negative values of $\beta$ for the second and third set of DAs, while it
differs about $50\%$ at $\beta=-1.4$ and $15\%$ at $\beta=-5$, for the first
set of DAs. For the positive values of $\beta$, $G_A(Q^2)$ practically
coincide for all three sets of DAs.    

\item The values of $G_P(Q^2)$ are very close to each other for all three
sets of DAs. 

\item The values of $G_A^S(Q^2)$ are very close to each other for the second
and third sets of DAs, but it is larger about $50\%$ for the first set of DAs
at $\beta=-1.4$.  

\item The difference between the values of $G_P^S(Q^2)$ is very small for
all three sets.

\item We see that all form factors are negative (positive) at positive
(negative) values of $\beta$.
   
\end{itemize}

For illustration, in Figs. (1)--(4) we present the dependencies of the form
factors $G_A(Q^2)$, $G_P(Q^2)$, $G_A^S(Q^2)$ and $G_P^S(Q^2)$ on $Q^2$ at
several fixed values of $\beta$, including the results for $\beta=-1$ case.
 
As has already been noted, the axial form factor $G_A(Q^2)$ is considered in
\cite{R8313}, however the analytical results that are presented in
that work and in ours should be different, since the considered structures are
different. However, we see that our numerical results on $G_A(Q^2)$ are close 
to that obtained in \cite{R8313} at $\beta=-1$.    

Note that for the analysis of the existing data from neutrino scattering
experiments, form factor $G_A(Q^2)$ is usually parametrized in the dipole
form 
\bea
G_A^{(d)}(Q^2) = \frac{g_A}{\ga 1+Q^2/m_A^2 \dr^2}~, \nnb
\eea
where $g_A=1.2695 \mp 0.0029$, which is determined from $\beta$ decay
\cite{R8302}. The global average for $m_A$ extracted from neutrino
scattering experiment is predicted to have the value 
$m_A=(1.026 \mp 0.021)~GeV$ in \cite{R8322}, while the value
$m_A=(1.20 \mp 0.12)~GeV$ announced by the K2K collaboration
is slightly larger \cite{R8323}.

In Figs. (5)--(10) we present the LCSR prediction for the axial form factor
$G_A(Q^2)$ normalized to $G_A^{(d)}(Q^2)$, for all three sets of DAs.
Experimentally this ratio, i.e., $G_A(Q^2)/G_A^{(d)}(Q^2)$ should be close
to $1$. From these figures we see that the prediction for this ratio by LCSR
is quite close to 1 at $\beta=-1.4$ for the first (third) set of DAs and for 
the  choice of the mass $m_A=1.0~GeV$ ($m_A=1.2~GeV$).
The ratio $R(Q^2) = G_A(Q^2)/G_A^d(Q^2)$ is close to 1 for the second set of
DAs at $\beta=-1$ for the choice of the mass $m_A=1.2~GeV$. But, as has
already been noted, $\beta$ should be different from $-1$. Therefore we can
conclude that the dipole form of the $G_A(Q^2)$ describes well only the
first set of DAs at $\beta=-1$. Our results on $G_A(Q^2)$ are in close 
agreement with the chiral quark model \cite{R8324} predictions and lattice 
results \cite{R8325} at $\beta=-1.4$. The results for $G_A(Q^2)$ for other 
values of $\beta$ depart considerably from the lattice results.
  
It should be remembered that all available experimental data exists
only at low $Q^2$ region. But unfortunately, our approach can give reliable
prediction about the form factors only at high $Q^2\ge 2~GeV^2$ region.
Therefore direct comparison of our theoretical prediction on experimental
result is impossible. Recently proposed experiment Minerva \cite{R8326} 
would provide a
precise determination of $G_A$ at $Q^2 < 2 ~GeV^2$ and $Q^2 > 2 ~GeV^2$.
Another experiment using $e+p \rar \nu+n$ has been planned at JLAB
\cite{R8327},
covering the range $Q^2 = 1 \div 3~GeV^2$. When the result of these experiments
will become available, it will be possible to compare the theoretical and
experimental results. As the final remark we would like to note that, 
for a more reliable prediction of the form factors, it is necessary to take 
into account the radiative $\alpha_s$ corrections and the distribution 
amplitudes with four, five particles.  

In conclusion, in this work we calculate the form factors of nucleons
induced by isovector and isoscalar axial--vector currents in LCSR method
using the general form of the nucleon interpolating current. In our
calculations we use three different sets. The dependence of the form
factor $G_A(Q^2)$ on $Q^2$ is compared with its dipole form parametrization
which follows from the analysis of neutrino experiments. Our analysis show
the results for $G_A(Q^2)$ in LCSR, for the first set of DAs at
$\beta=-1.4$, well describes the dipole form of $G_A(Q^2)$.

\section*{Acknowledgments}

One of the authors (T. M. A) is grateful to T\"{U}B\.{I}TAK for partially
support of this work under the project 105T131.

\newpage

\section*{Figure captions}
{\bf Fig. (1)} The dependence of the nucleon form factor $G_A(Q^2)$ 
on $Q^2$ at $M^2=2~GeV^2$ and $s_0=2.25~GeV^2$, 
at four different values of $\beta$: $\beta=-5$, $\beta=-1.4$ $\beta=-1$ 
and $\beta=5$, for the first set of DAs. \\ \\
{\bf Fig. (2)} The same as in Fig. (1), but for the form factor
$G_P(Q^2)$.\\ \\
{\bf Fig. (3)} The same as in Fig. (1), but for the form factor
$G_A^S(Q^2)$. \\ \\
{\bf Fig. (4)} The same as in Fig. (1), but for the form factor
$G_P^S(Q^2)$.\\ \\
{\bf Fig. (5)} The dependence of the ratio $R=G_A(Q^2)/G_A^d(Q^2)$ on
$Q^2$ at three fixed values of $\beta$, for the first set of DAs at
$m_A=1.0~GeV$.\\ \\
{\bf Fig. (6)} The same as in Fig. (5), but for $m_A=1.2~GeV$.\\ \\
{\bf Fig. (7)} The same as in Fig. (5), but for the second set of DAs.\\ \\
{\bf Fig. (8)} The same as in Fig. (6), but for the second set of DAs.\\ \\
{\bf Fig. (9)} The same as in Fig. (5), but for the third set of DAs.\\ \\
{\bf Fig. (10)} The same as in Fig. (6), but for the third set of DAs.

\newpage

\begin{figure}
\vskip 3. cm
    \includegraphics{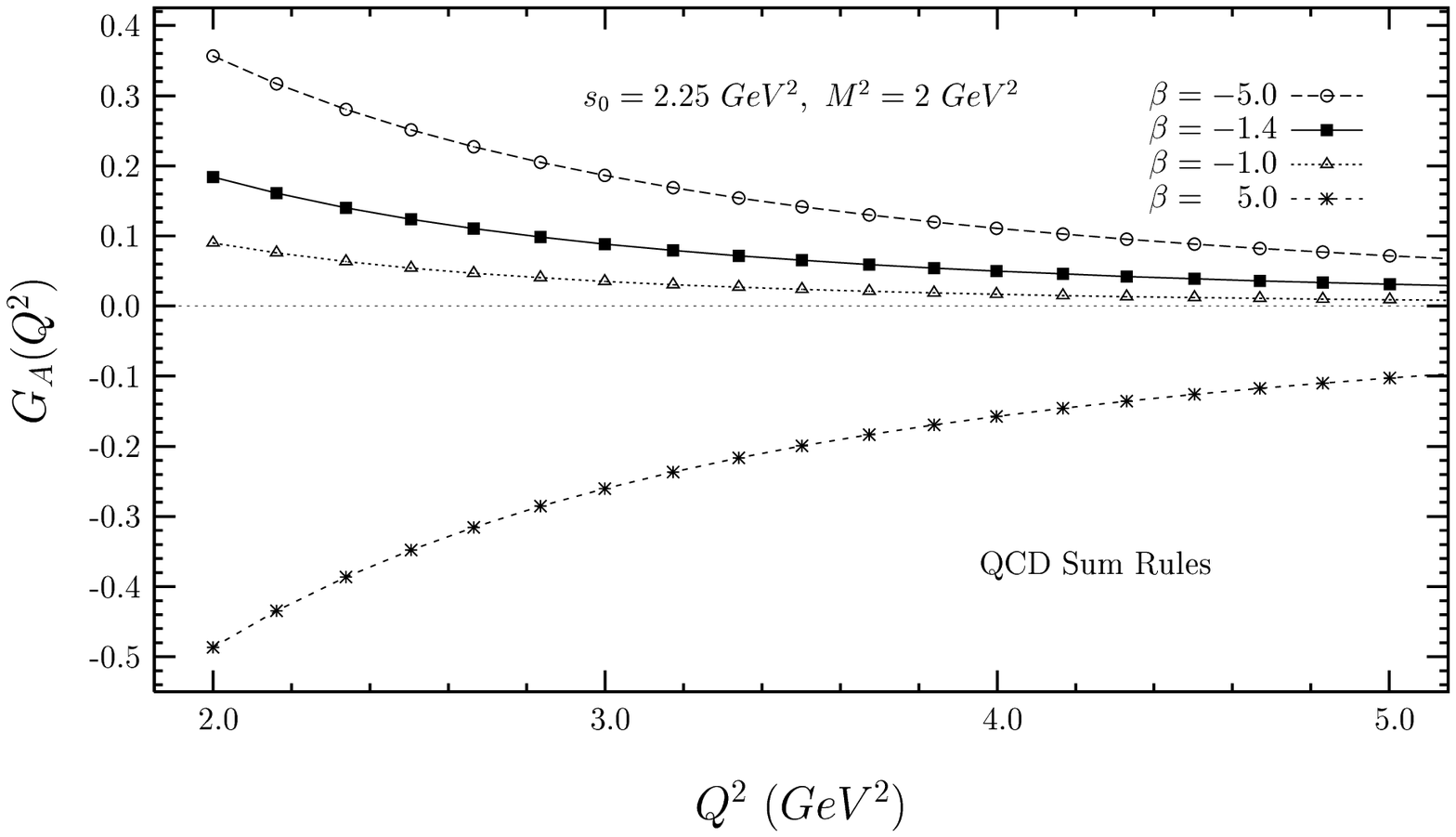}
\vskip 6.3cm
\caption{}
\end{figure}

\begin{figure}
\vskip 4.0 cm
    \includegraphics{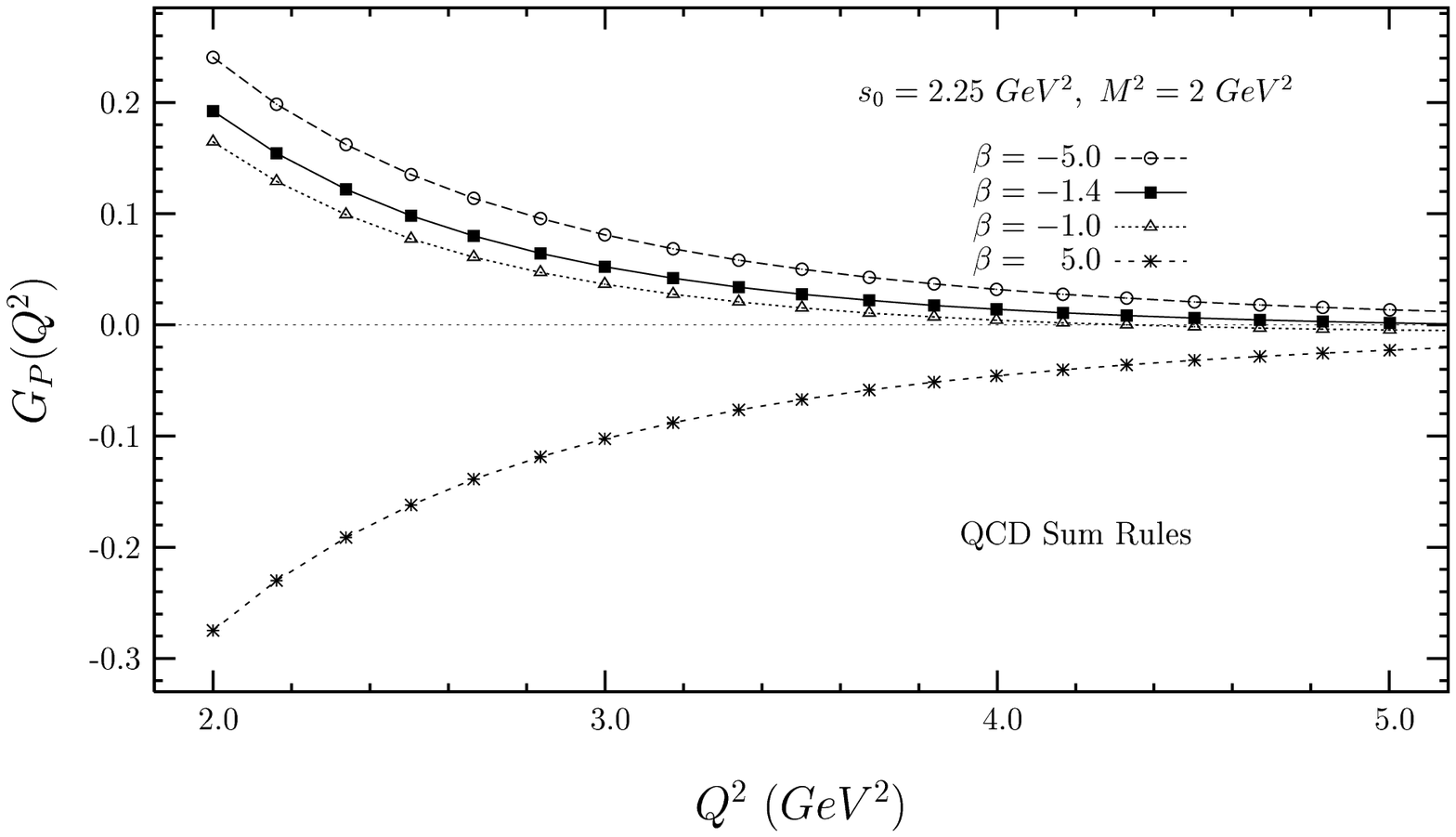}
\vskip 6.3 cm
\caption{}
\end{figure}

\newpage

\begin{figure}
\vskip 3. cm
    \includegraphics{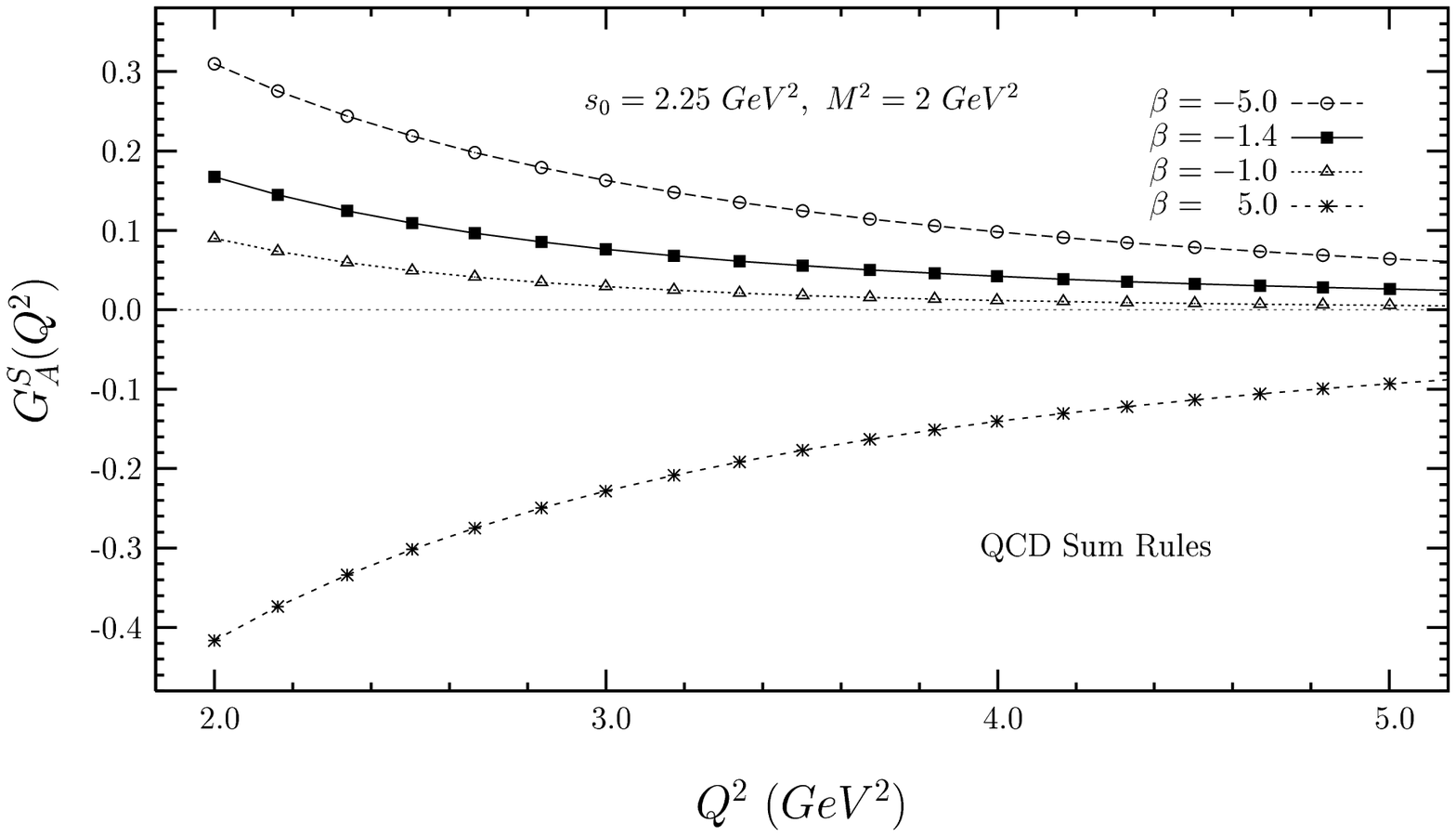}
\vskip 6.3cm
\caption{}
\end{figure}

\begin{figure}
\vskip 4.0 cm
    \includegraphics{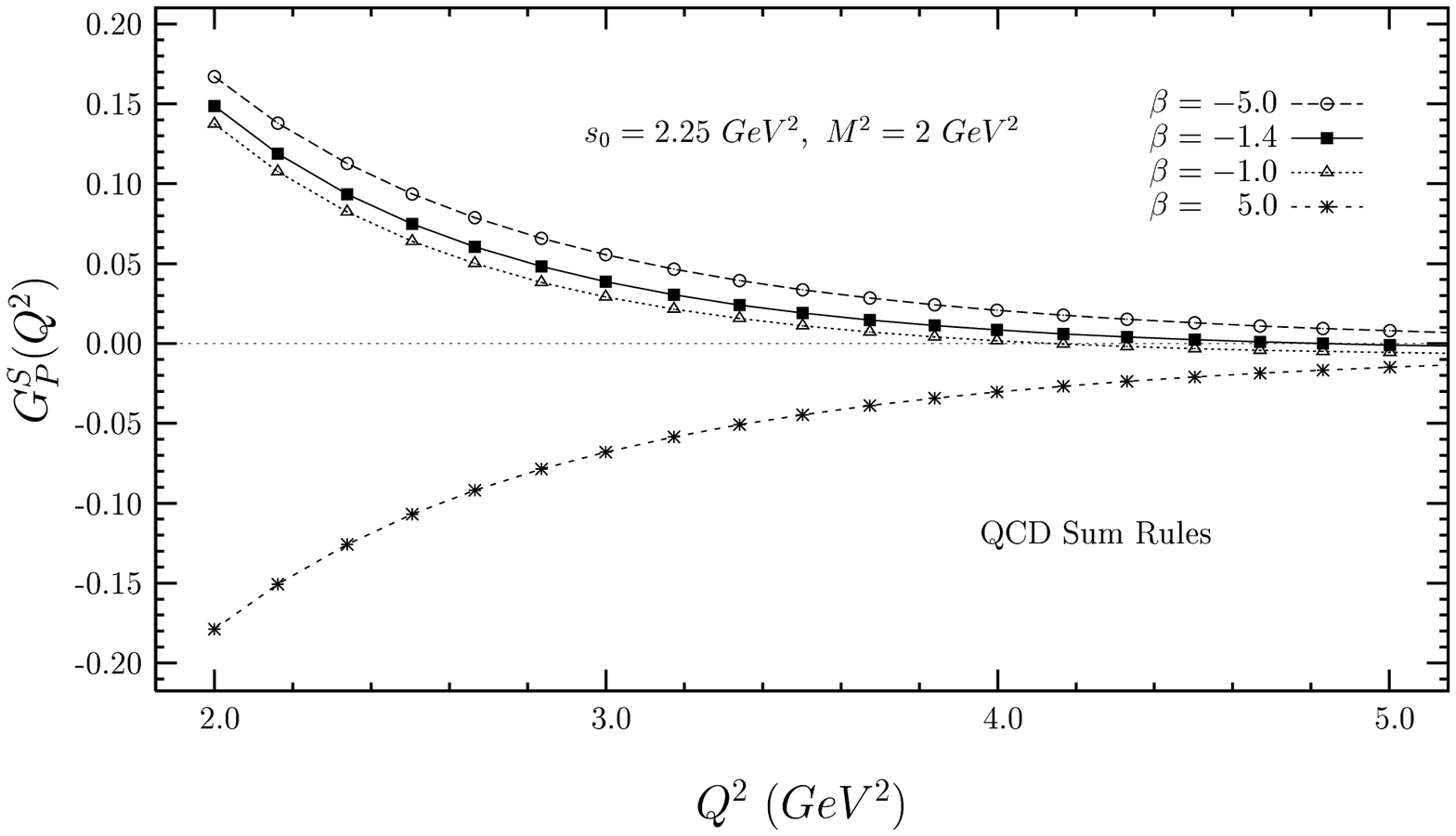}
\vskip 6.3 cm
\caption{}
\end{figure}

\begin{figure}
\vskip 3. cm
    \includegraphics{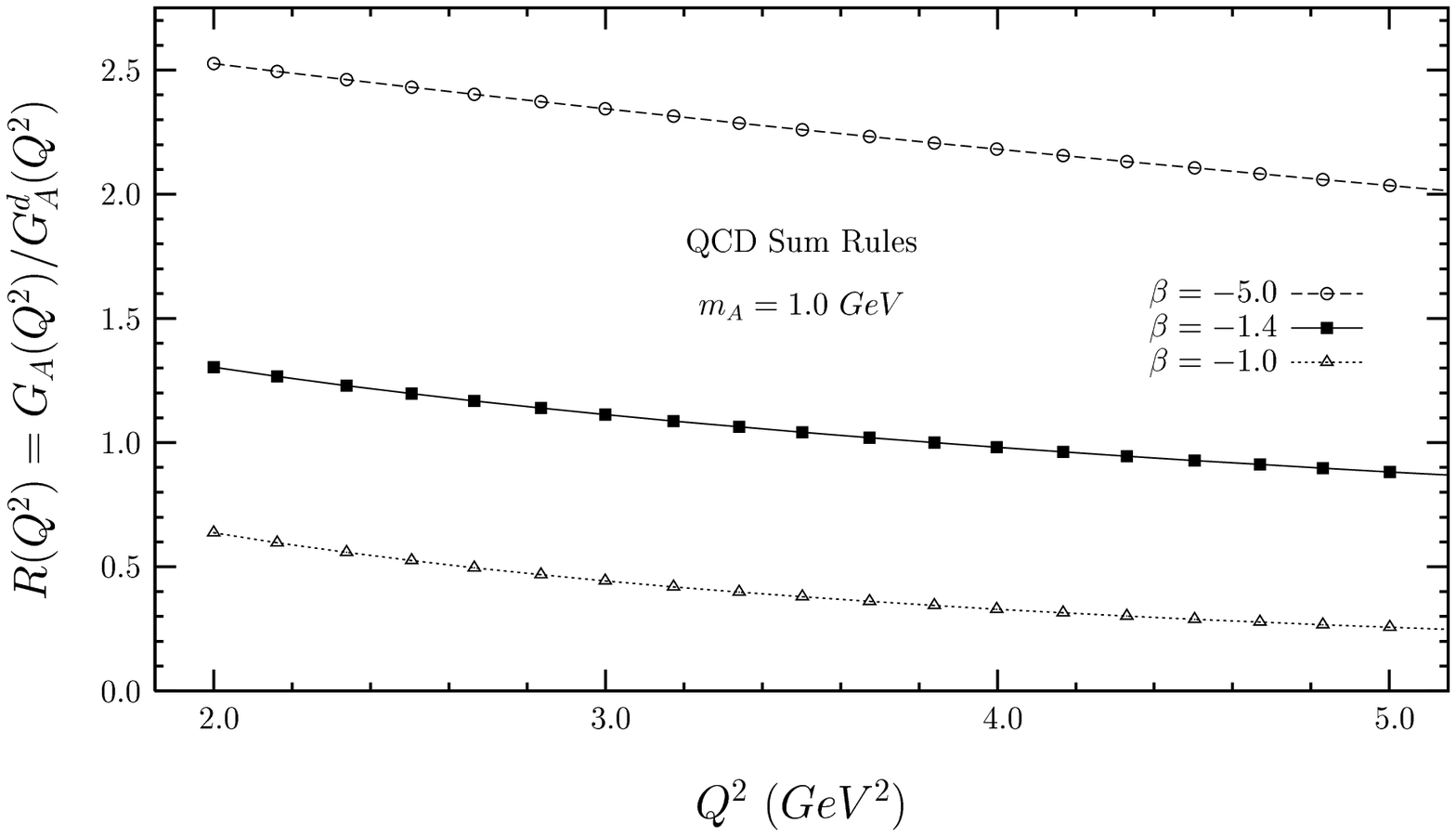}
\vskip 6.3cm
\caption{}
\end{figure}

\begin{figure}
\vskip 4.0 cm
    \includegraphics{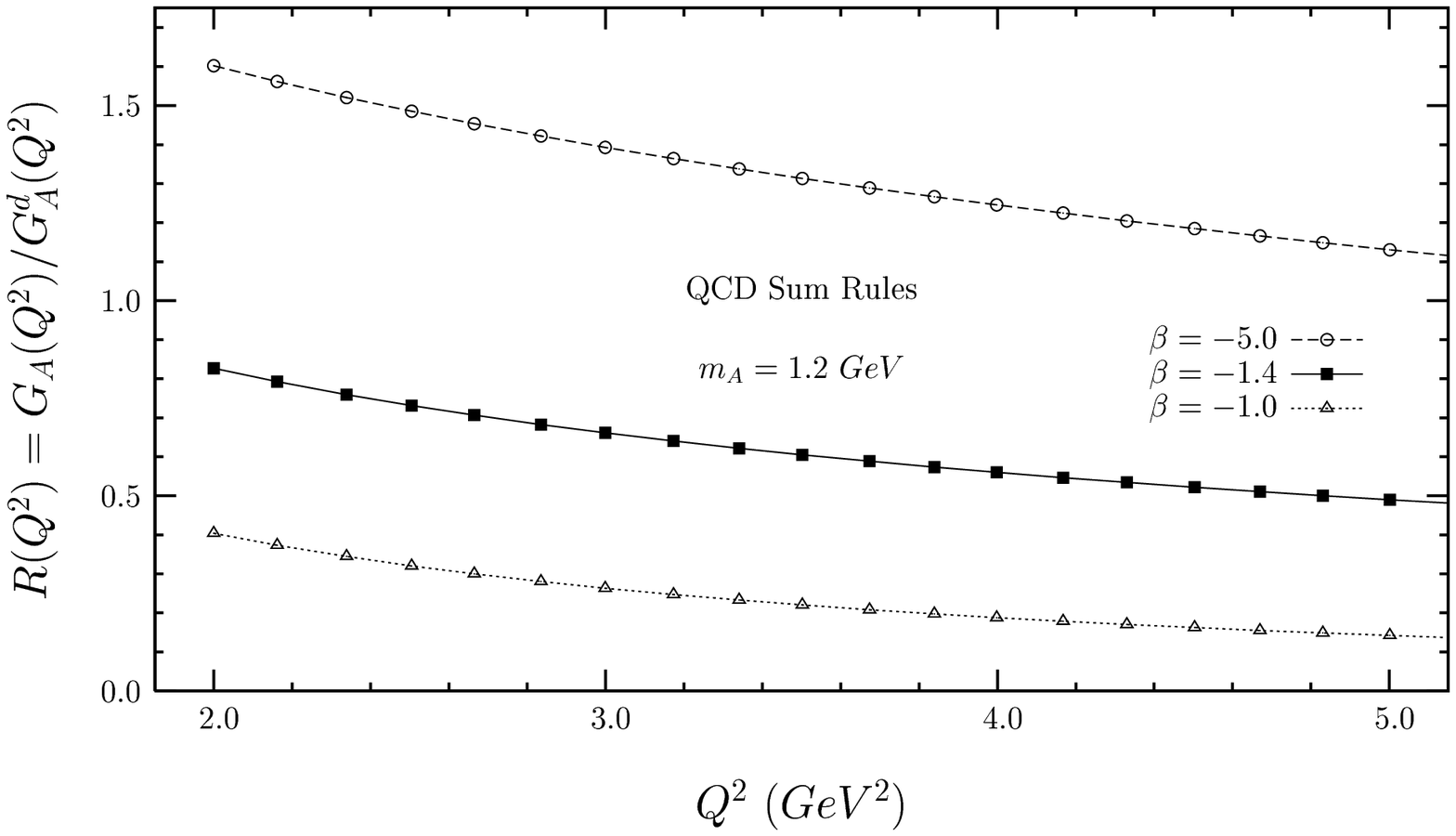}
\vskip 6.3 cm
\caption{}
\end{figure}

\newpage

\begin{figure}
\vskip 3. cm
    \includegraphics{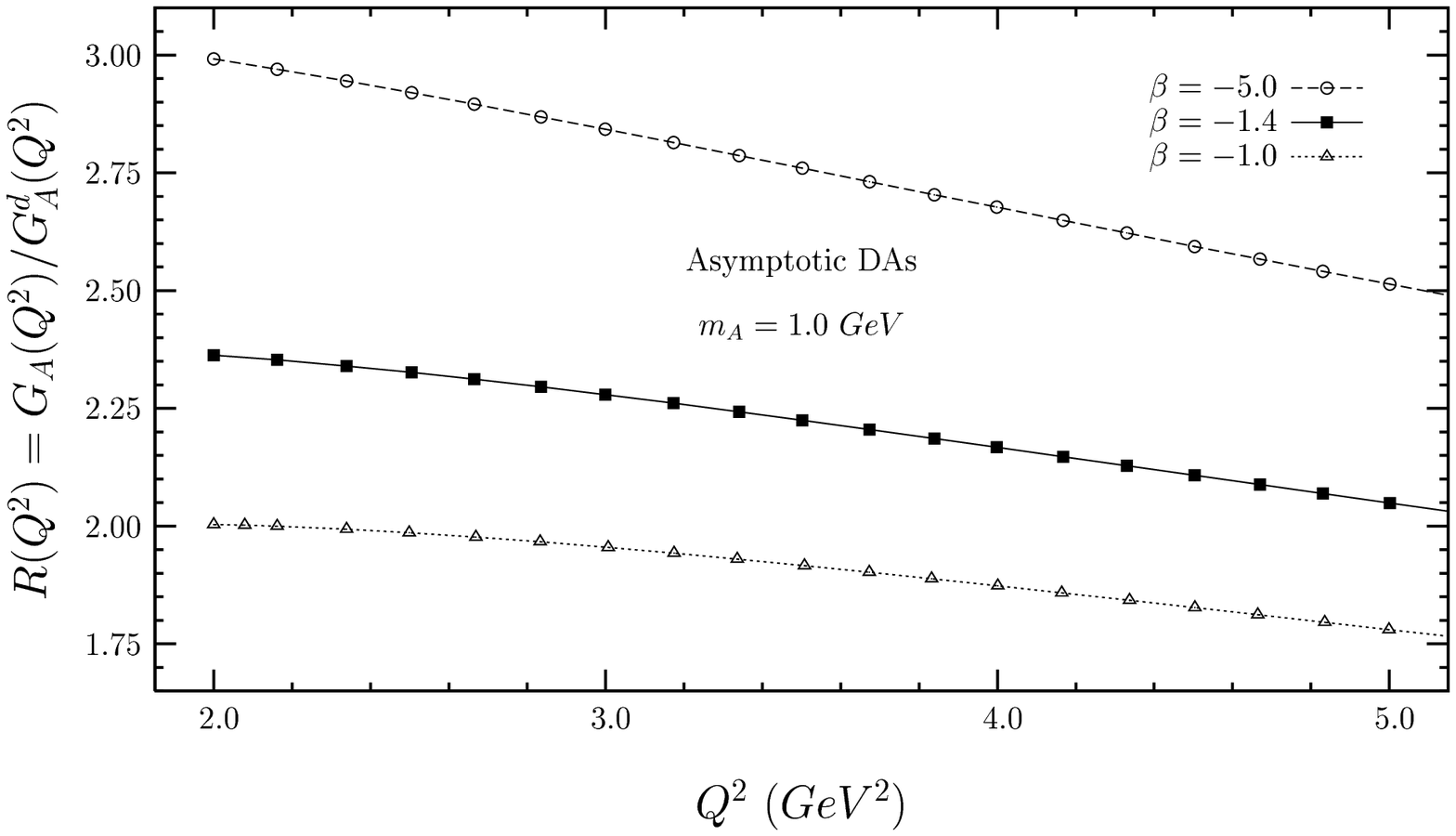}
\vskip 6.3cm
\caption{}
\end{figure}

\begin{figure}
\vskip 4.0 cm
    \includegraphics{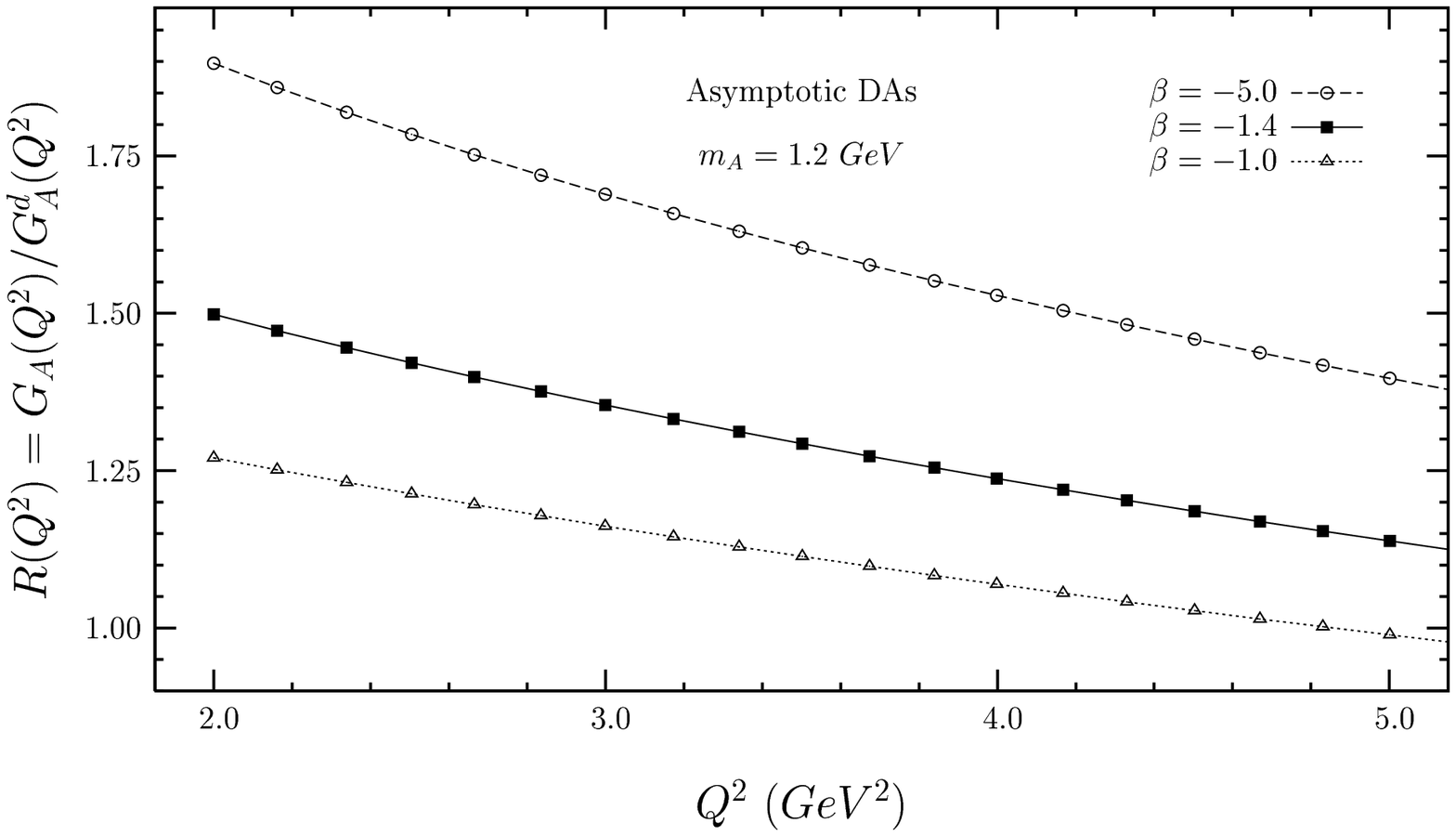}
\vskip 6.3 cm
\caption{}
\end{figure}

\newpage

\begin{figure}
\vskip 3. cm
    \includegraphics{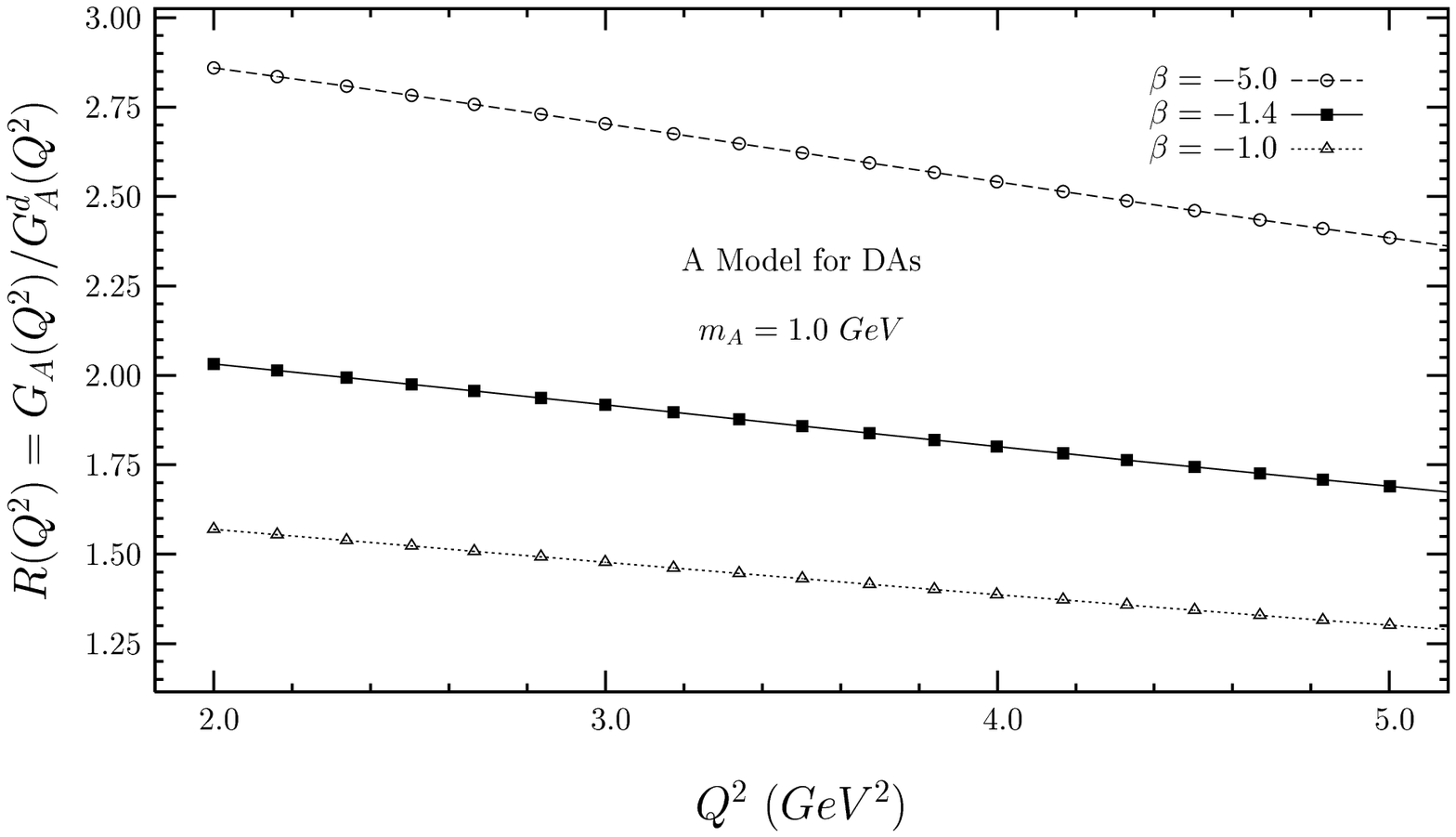}
\vskip 6.3cm
\caption{}
\end{figure}

\begin{figure}
\vskip 4.0 cm
    \includegraphics{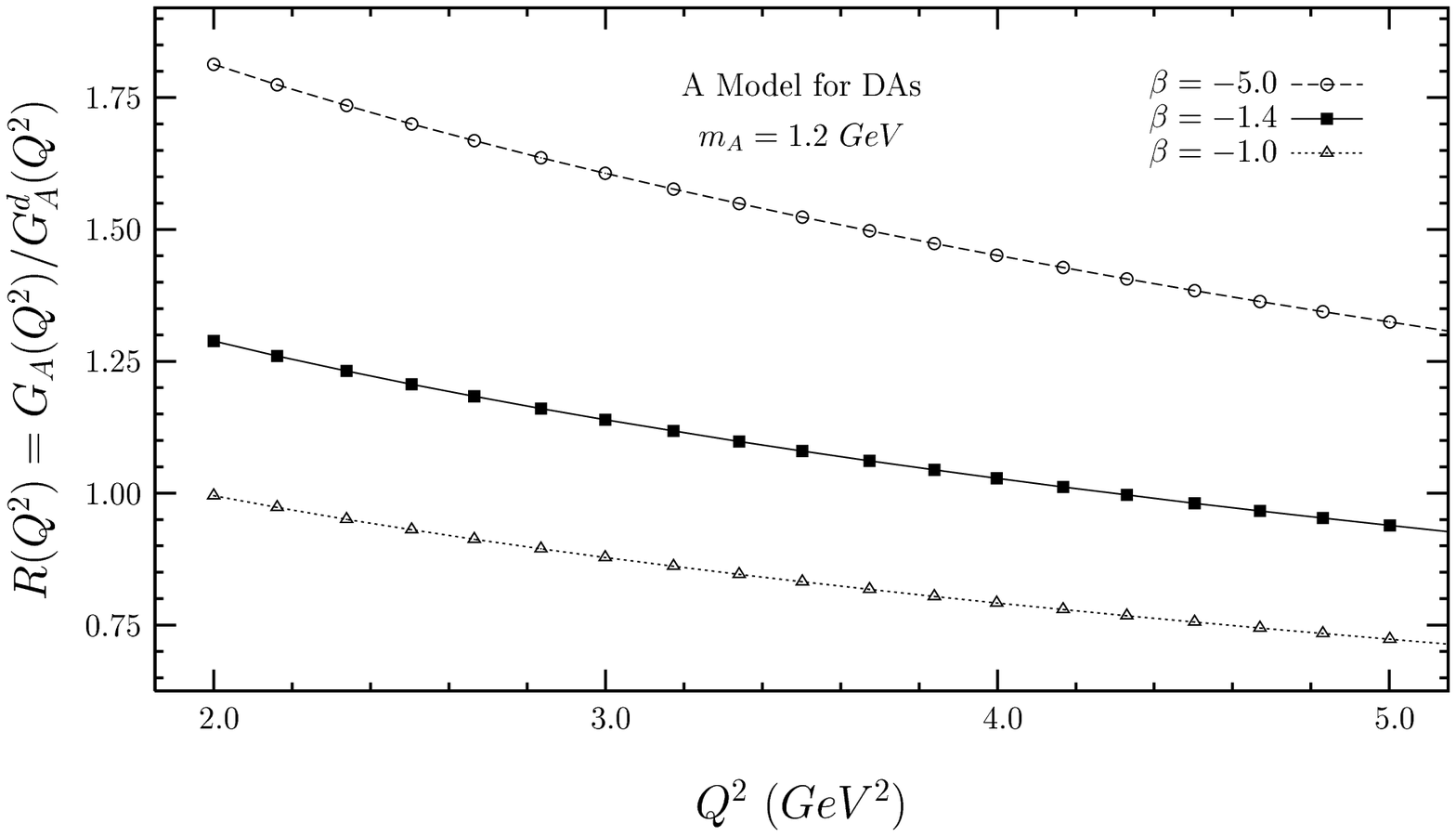}
\vskip 6.3 cm
\caption{}
\end{figure}

\end{document}